\documentclass[11pt]{article} %
\usepackage{pgfplots}
\pgfplotsset{width=7cm,compat=1.9}
\usepackage{authblk}
\usepackage{booktabs}
\usepackage{float}
\usepackage[margin=1in]{geometry}
\usepackage[scr=rsfso]{mathalfa}
\usepackage{mdframed} %
\usepackage{microtype} %
\usepackage{mathpazo}
\usepackage{amsthm}
\usepackage{xspace}
\usepackage{comment}
\usepackage{lipsum}
\usepackage{amsfonts}
\usepackage{graphicx}
\usepackage{epstopdf}
\usepackage{algorithm}
\usepackage[noend]{algorithmic}
\usepackage{tikz}
\usepackage{caption}
\usepackage{subcaption}
\usepackage{amsmath}
\usepackage{hyperref}
\usepackage{amssymb}
\usepackage{dirtytalk}
\usepackage{mathtools}
\usepackage{xspace}
\usepackage[colorinlistoftodos,textsize=tiny,textwidth=2cm,color=green!50!gray]{todonotes}
\usetikzlibrary{decorations.pathmorphing}
\ifpdf
  \DeclareGraphicsExtensions{.eps,.pdf,.png,.jpg}
\else
  \DeclareGraphicsExtensions{.eps}
\fi

\let\epsilon=\varepsilon %

\newcommand{\flmi}{\textsc{FL-MI}\xspace}

\newcommand{\VC}{V_C}

\newcommand{\cost}{\operatorname{cost}}
\newcommand{\dist}{\operatorname{dist}}
\newcommand{\mc}{\operatorname{mc}}

\newtheorem{theorem}{Theorem} 
\newtheorem{lemma}[theorem]{Lemma} %
\newtheorem{definition}[theorem]{Definition} %
\newtheorem{conjecture}[theorem]{Conjecture} %

\usepackage[inline]{enumitem}
\setlist[enumerate,1]{label=(\roman*)}
\setlist[enumerate,2]{label=(\alph*)}

\newlist{stepsroman}{enumerate}{1}
	\setlist[stepsroman]{label=(\roman*)}

\newlist{stepsarabic}{enumerate}{1}
	\setlist[stepsarabic]{label=\arabic*., ref=\arabic*}

\newlist{stepsalph}{enumerate}{1}
	\setlist[stepsalph]{label=(\alph*), ref=(\alph*)}

\usepackage[capitalize, nameinlink]{cleveref}
\crefname{theorem}{Theorem}{Theorems}
\Crefname{lemma}{Lemma}{Lemmas}
\Crefname{claim}{Claim}{Claims}
\Crefname{fact}{Fact}{Facts}
\Crefname{remark}{Remark}{Remarks}
\Crefname{observation}{Observation}{Observations}
\Crefname{figure}{Figure}{Figures}
\Crefname{line}{Line}{Lines}
\Crefname{algocf}{Algorithm}{Algorithms}
\crefalias{AlgoLine}{line}
\Crefname{stepsromani}{Step}{Steps}
\Crefname{stepsarabici}{Step}{Steps}
\Crefname{stepsalph}{Step}{Steps}

\AddToHook{env/lemma/begin}{\crefalias{theorem}{lemma}}
\AddToHook{env/claim/begin}{\crefalias{theorem}{claim}}
\AddToHook{env/observation/begin}{\crefalias{theorem}{observation}}
\AddToHook{env/definition/begin}{\crefalias{theorem}{definition}}

\def\DEBUG{true} 
\ifdefined\DEBUG

\newcommand{\ramr}[1]{\todo[color=blue!100!white!50]{R: #1}}
\newcommand{\zacr}[1]{\todo[color=orange!100!black!50]{Z: #1}}
\newcommand{\fabr}[1]{\todo[color=red!100!black!50]{F: #1}}
\newcommand{\mrsr}[1]{\todo[color=green!100!black!50]{M: #1}}
\newcommand{\kinr}[1]{\todo[color=cyan!100!black!50]{K: #1}}
\else

\newcommand{\ramr}[1]{}

\newcommand{\fabr}[1]{}

\newcommand{\zacr}[1]{}

\newcommand{\mrsr}[1]{}

\newcommand{\kinr}[1]{}

\fi

\usepackage[
backend=biber,
style=alphabetic,
citestyle=alphabetic,
maxalphanames=6,
maxcitenames=99,
mincitenames=98,
maxbibnames=99,
giveninits=true,
url=false,
doi=true,
isbn=true,
backref=true
]{biblatex}

\begin{document}

\title{Facility Location with Customer Incompatibilities}
\author[1]{Zachary Friggstad}
\author[1]{Pengfei Wu}
\affil[1]{Department of Computing Science, University of Alberta}

\date{}

\maketitle

\thispagestyle{empty}
\begin{abstract}
We study metric facility location with customer incompatibilities, where the clients assigned to each facility must satisfy prescribed independence constraints. We first consider the setting of  facility-opening costs and where there is a single matroid $\mathcal M$ over the clients and the set of clients assigned to an open facility must be an independent set in $\mathcal M$. We develop a polynomial-time \((1+\sqrt{2}+\epsilon)\)-approximation algorithm for this setting. We further obtain a polynomial-time \((15+\epsilon)\)-approximation algorithm for the case of non-uniform facility-opening costs when $\mathcal M$ is a partition matroid that only permits $b_g$ clients to be assigned to a facility for each group $g$ in the partition. Both settings are substantial generalizations of Capacitated Facility Location with uniform capacities.

Our algorithms are based on local search using add, delete, and single-swap operations. As with CFL, bounding the connection cost is reasonably straightforward using supermodularity of the minimum connection-cost function.
However, bounding the facility opening costs requires novel arguments in both cases. 
We use a ``compression'' argument based on matroid exchange principles to identify a subset of the locally opened facilities whose size is no larger than that of an optimal facility set. The opening costs of facilities outside this subset can be bounded using supermodularity, while bounding the costs of the remaining facilities requires different arguments.

It is natural to consider a variant where each possible facility $i$ has its own matroid $\mathcal M_i$ defining the subsets of clients it can feasibly serve. We also show problem is at least as hard to approximate as Set Cover even if all $\mathcal M_i$ are partition matroids that have the same partition of clients, just different group bounds $b_g$.


\end{abstract}

\newpage
\setcounter{page}{1}

\section{Introduction}
Facility location is a classical problem in combinatorial optimization, with applications to distribution-system design and customer allocation~\cite{kloseDrexl2005}. In the approximation-algorithms literature, several classical variants have been extensively studied, including
Uncapacitated Facility Location~\cite{Li2013UFL},
$k$-Median~\cite{AryaEtAl2004},
Capacitated Facility Location~\cite{AnSinghSvensson2017},
Capacitated $k$-median~\cite{DemirciLi2016},
Capacitated Facility Location with uniform capacities
~\cite{AggarwalEtAl2013},
and Capacitated Facility Location with uniform opening costs
~\cite{Kao2023}.

Customer incompatibilities have been studied in the operations-research
literature under the name \emph{Simple Plant Location Problem with
Incompatibilities}. Mar\'in and Pelegr\'in~\cite{marinPelegrin2019}
studied a model in which specified pairs of clients cannot be served by
the same facility. They developed a set-packing formulation, derived
facet-defining inequalities, and evaluated separation procedures in a
computational study. Pelegr\'in~\cite{pelegrin2023} subsequently revisited
this model together with co-location variants and discussed applications
in telecommunication network design. Related capacitated variants with
customer incompatibilities have also been studied using metaheuristics,
including multi-neighborhood simulated
annealing~\cite{maiaEtAl2023,ceschiaSchaerf2024}.

We study metric Facility Location with Customer Incompatibilities, where
additional constraints restrict which clients may be served by the
same facility. The objective is to minimize the total facility-opening
and connection costs while assigning every client to an open facility
and respecting these constraints.

The model of customer incompatibilities where we specify pairs of clients that cannot be assigned to the same facility is clearly a very hard problem to approximate since we could only schedule independent sets from the incompatibility graph at facilities, we make this hardness more formal in Appendix \label{app:overlapping-groups-hardness}. So we consider other incompatibility notions in this paper that admit efficient constant-factor approximations.

Specifically, the core problem we study is the following.
\begin{definition}
In {\em Facility Location with Matroid-based Incompatibilities} (\flmi), we are given a set of clients $V_C$, a set of facilities $V_F$, metric distances $d(i,j)$ for $i,j \in V_C \cup V_F$, and opening costs $f_i \geq 0$ for $i \in V_F$.
Additionally, we are given a matroid $\mathcal M = (V_C, \mathcal I)$ describing feasible assignments of clients to a facility.

Given $S \subseteq V_F$, we say an assignment $\pi : V_C \rightarrow S$ is {\em feasible} if $\pi^{-1}(i) \in \mathcal I$ for each $i \in S$. We similarly say $S \subseteq V_F$ is {\em feasible} if there is a feasible assignment $\pi : V_C \rightarrow S$.

The goal is to open a feasible subset $S \subseteq V_F$ and find a feasible $\pi : V_C \rightarrow S$ minimizing the opening cost and connection cost: $\sum_{i \in S} f_i + \sum_{j \in V_C} d(j, \pi(j))$.
\end{definition}

When the matroid is a uniform matroid, this is exactly the Capacitated Facility Location problem with uniform capacities. Other matroids capture settings of interest in operations resesarch.
\begin{itemize}
    \item {\bf Transversal matroids}: Here, at each facility there is also a set of resources $R$. Each client $j$ has a subset $R_j \subseteq R$ of resources that could be used to serve it. A subset of clients $X \subseteq V_C$ can then be served at a facility if there is a one-to-one mapping $\rho : X \rightarrow R$ such that $\rho(j) \in R_j$ for each client $j \in X$, i.e. the clients in $X$ can be matched to resources.
    \item {\bf Laminar matroids}: Here we have a laminar family $\mathcal L$ over $V_C$, each $A \in \mathcal L$ has a bound $b_A \geq 0$. A subset $X$ of clients can be assigned to a facility only if $|X \cap A| \leq b_A$ for each $A \in \mathcal L$. This is a significant generalization of Capacitated Facility Location with uniform capacities; it captures a more hierarchical structure of constraints beyond simply bounding the total number of clients being served.
\end{itemize}

Throughout this paper, we assume the matroids are given via efficient independence oracles: given a subset $X \subseteq V_C$ we are able to tell if $X \in \mathcal I$ in time that is polynomial in $|V_C|$ and, perhaps, other additional structures given to present the matroid. Essentially, our algorithms make a polynomial number of calls to independence oracles and all other steps run in time that is polynomial in $V_C, V_F$ and the number of bits required to encode rational distances $d$. We also assume the whole set $V_F$ itself is feasible, as otherwise no feasible solution could exist.

We remark that given $S \subseteq V_F$, we can determine if it is feasible and, if so, find a minimum-cost feasible assignment $\pi : V_C \rightarrow S$ in polynomial time by computing a minimum-cost way to partition $V_C$ into $|S|$ independent sets, assigning a cost of $d(j, \pi(j))$ for
assigning $j$ to the independent set corresponding to facility $\sigma(j)$. We say an instance of \flmi has {\em uniform opening costs} if all opening costs $f_i, i \in V_F$ are the same value $f$. We sometimes use the term {\em non-uniform opening costs} to emphasize when we are not specializing to the uniform opening
cost setting \cite{schrijver2003}.

Finally, we also focus the following important special case.
\begin{definition}
\flmi with {\em group constraints} is the case where $\mathcal M$ is a partition matroid. That is, the clients are partitioned into groups, each group $g$ has a bound $b_g \geq 0$, and a subset $X \subseteq V_C$ is independent if for each group $g$ we have at most $b_g$ clients from group $g$ in $X$.
\end{definition}
Note \flmi with group constraints is still a substantial generalization of Capacitated Facility Location with uniform capacities, since that is the case where the partition matroid has only one group (i.e. is a uniform matroid).

The partition assumption is a key structural restriction of our
group-constrained model. If clients are allowed to belong to
multiple groups without further restrictions, the problem already
captures Graph Coloring through a simple approximation-preserving
reduction (see Appendix~\ref{app:overlapping-groups-hardness}).
Consequently, the more general problem with overlapping groups of clients admits no polynomial-time
$n^{1-\epsilon}$-approximation for any fixed $\epsilon>0$
unless $P=NP$, where $n$ is the number of clients
\cite{zuckerman2007}.

The requirement that all facilities share the \emph{same} client
matroid is a similarly important structural restriction of our
matroid-constrained model. Allowing facility-specific client
matroids already captures Set Cover, even with uniform opening
costs and zero connection costs, through a simple
approximation-preserving reduction (see
Appendix~\ref{app:facility-specific-hardness}).
Consequently, the problem admits no polynomial-time
$(1-\eta)\ln n$-approximation for any fixed $0<\eta<1$
unless $P=NP$, where $n$ is the number of clients
\cite{DinurSteurerSetCover2014}.
In particular, no polynomial-time constant-factor approximation
exists unless $P=NP$.

\subsection{Related Work}
\label{subsec:related-work}

\paragraph{Capacitated Facility Location.}
The capacitated facility location problem (CFL) is a
well-studied generalization of uncapacitated facility
location (UFL), for which Li~\cite{Li2013UFL}
obtained a $1.488$-approximation.
In CFL, each facility has a capacity limiting
the number of clients it can serve.
For general capacities and opening costs,
Bansal et al.~\cite{BansalGargGupta2012}
gave a $5$-approximation using local search.
An et al.~\cite{AnSinghSvensson2017}
developed the first LP-based constant-factor
approximation for general CFL by strengthening
the standard LP relaxation.

Two important special cases are CFL with uniform
capacities and CFL with uniform opening costs.
For uniform capacities, Aggarwal et
al.~\cite{AggarwalEtAl2013} obtained a
$3$-approximation, even with non-uniform
opening costs.
For uniform opening costs, Kao~\cite{Kao2023}
obtained a $4$-approximation, which was
subsequently improved to a
$(3.733+\epsilon)$-approximation by Gupta et
al.~\cite{GuptaDabasGarg2026} using local search.
Both variants are special cases of our model:
uniform capacities correspond to a partition
matroid with a single group, while our general
matroid constraints allow more complicated
restrictions on client assignments.

\paragraph{Matroid Median and Capacitated $k$-Median.}
Another related problem is matroid median, which generalizes
$k$-median by requiring the set of opened facilities to be
independent in a given matroid.
Krishnaswamy et al.~\cite{KrishnaswamyEtAl2011} introduced
this problem and obtained a $16$-approximation.
Swamy~\cite{Swamy2016} improved the guarantee to $8$,
and Krishnaswamy et al.~\cite{KrishnaswamyLiSandeep2018}
subsequently obtained a $7.081$-approximation
using iterative rounding.
Unlike our model, where the matroid restricts the clients
assigned to each facility, matroid median imposes
constraints on the set of opened facilities.

A natural extension of our problem would be to impose
matroid constraints on both the opened facilities and
the client assignments.
This extension already captures uniform capacitated
$k$-median, where at most $k$ facilities may be opened
and each facility may serve at most $u$ clients.
These two restrictions correspond to rank-$k$ and
rank-$u$ uniform matroids on the facility and client
sets, respectively, with zero facility-opening costs.
Despite progress on constant-factor approximations
allowing $(1+\epsilon)$ violations of either the
facility bound or the capacities
~\cite{Li2015UniformCKM,DemirciLi2016},
no polynomial-time constant-factor approximation
satisfying both constraints is known
~\cite{DaiLiPeng2026}.
Thus, extending our results to include matroid
constraints on opened facilities would already
encompass this longstanding open problem,
even when both matroids are uniform.

\subsection{Our Results and Techniques}\label{subsec:our-results-and-techniques}


Our algorithms are based on local search over sets of open facilities.
We consider the following three local search operations. Given a feasible $S \subseteq V_F$, we consider the following operations:
\begin{itemize}
\item \textbf{add($i$)}: For some $i\in V_F\setminus S$, replace $S$ by
$S\cup\{i\}$.
\item \textbf{delete($i$)}: For some $i\in S$, replace $S$ by
$S\setminus\{i\}$, provided the resulting facility set is feasible.
\item \textbf{swap($s,i$)}: For some $s\in S$ and $i\in V_F\setminus S$,
replace $S$ by $(S\setminus\{s\})\cup\{i\}$.
\end{itemize}

Let $\mathcal N(S)$ be all possible feasible subsets of $S$ that can be obtained by applying one of these operations.

\begin{algorithm}
\caption{Local Search for \flmi}
\begin{algorithmic}
\STATE $S \gets V_F$
\WHILE{$\cost(S') < \cost(S)$ for some $S' \in \mathcal N(S)$}
    \STATE $S' \gets S$
\ENDWHILE
\STATE \textbf{return} $S$
\end{algorithmic}
\end{algorithm}
In the uniform-opening-cost setting, we only need to consider add and delete moves to define $\mathcal N(S)$;
for non-uniform opening costs, we also use swap moves. Say $S$ is a {\em local minimum} if every $S" \in \mathcal N(S)$ has $\cost(S') \geq \cost(S)$,
i.e. the algorithm would stop if it reached this set $S$.

Each iteration can be executed in polynomial time because $|\mathcal N(S)| = O(|V_F|^2)$ and calculating $\cost(S')$ can be done efficiently.
Our analysis will only use $2|V_F|$ test swaps, so the standard technique of only updating $S' \gets S$ if $\cost(S') \leq (1 - \epsilon/(2|V_f|)) \cost(S)$
applies to show that the approximation guarantee of the final set $S'$ is at most a $(1+\epsilon)$-factor worse than the approximation guarantee for true
locally-minimum solutions $S$. But now the algorithms run in polynomial time (for rational distances and opening costs) for every constant $\epsilon > 0$ since $\cost(S)$ decreases by a constant factor every $2 |V_F|/\epsilon$ iterations.
We ignore this detail from now on and focus on analyzing the approximation guarantee of a locally-minimum solution.


\begin{theorem}
\label{thm:general-matroid-approximation}
There exists a \((1+\sqrt{2}+\epsilon)\)-approximation algorithm for \flmi with uniform opening costs.
\end{theorem}

Turning to non-uniform opening costs, we are able to prove that local search yields a good approximation in the case of group constraints.
\begin{theorem}
\label{thm:partition-matroid-nonuniform}
There exists a \((15+\epsilon)\)-approximation algorithm for \flmi with group constraints and non-uniform facility-opening costs.
\end{theorem}

A key ingredient in our analysis is the supermodularity of the
minimum connection-cost function $A(\cdot)$.
This property allows us to derive connection cost inequalities
from local optimality under add and delete moves, without
explicitly constructing feasible client reassignments for
each test move.

Bounding the facility opening cost is the more challenging task.
Letting $S$ denote a locally-optimum solution and $O$ a globally-optimum
solution, we first choose a feasible assignment $\pi:V_C\to S$
minimizing
   $$ W(\pi):=\sum_{j\in V_C}d\left(\sigma_O(j),\pi(j)\right).$$
We show that using appropriate tie-breaking criteria when some costs are equal,
the image $P := \pi(V_C)$ satisfies $|P| \leq |O|$ and, further, that $A(P) \leq A(S) + 2 \cdot A(O)$ where $A(X)$ is the minimum-cost of a feasible assignment $\sigma : V_C \rightarrow X$ using $A(X) = \infty$ if $X$ is not feasible.
Proving $|P| \leq |O|$ is subtle and uses properties similar to exchange paths in matroid intersection algorithms.

For the case of \flmi, supermodularity of $A(\cdot)$ and the above bound on $A(P)$ allow us to bound the opening cost $F(S\setminus P)$ by $2A(O)$
in another relatively straightforward manner.

For non-uniform opening costs, our bound on $F(S\setminus P)$ from the case of uniform opening costs essentially continues to hold.
However, the bound $|P| \leq |O|$  is no longer sufficient to control the total opening cost.
We therefore introduce swap moves to compare the opening costs
of facilities in $P$ with those in $O$.

However, this creates a new difficulty: we need to ensure each facility in $O$ is not swapped in too many time
to avoid charging its cost too many times in the analysis.
In the case of group constraints, we overcome this difficulty, by first adding $S \cap O$ to $P$ and then finding mapping $\nu : P \to O$ such that each $i \in O$ is mapped to at most $8$ times
to bound the number of times it is opened in the corresponding test swaps.
This mapping will be very carefully constructed so that
\[
    \sum_{s\in T}
    \min\{d(s,\nu(s)),\dist(s,P\setminus T)\}
    \le 6\mc(T,O)
\]
for every $T\subseteq P$,
where $\mc(T,O)$ is the minimum total distance of an injection from
$T$ to $O$, and $\dist(s,\emptyset):=+\infty$.
The simultaneous distance bound controls reassignment costs
in feasible test swaps and allows us to handle reassignments for each group separately.
Of course, care needs to be taken to complete the analysis, but this is the key contribution of our approach.




\section{Preliminaries and Common Results}\label{sec:notation}
In this section, we establish some notation used throughout the paper and establish supporting results that will be used to prove Theorems \ref{thm:general-matroid-approximation} and \ref{thm:partition-matroid-nonuniform}.

\subsection{Notation}
Recall that $A(X)$ for $X \subseteq V_F$ is the optimum connection cost over all feasible assignments $\sigma : V_C \rightarrow X$, using $A(X) = \infty$ if there is no feasible assignment. Also recall $F(X) = \sum_{i \in X} f_i$ is the opening cost of $X$.

We let $S \subseteq V_F$ be some fixed locally optimal solution and $O \subseteq V_F$ be some fixed globally optimal solution. For brevity, we use $C = A(S), F = F(S)$ denote the connection and opening costs of $S$
and $C^* = A(O), F^* = F(O)$ be the connection and opening costs of $O$. We also let $\sigma_S : V_C \rightarrow S$ and $\sigma_O : V_C \rightarrow O$ be optimal feasible assignments for the respective sets of facilities $S,O$.

For any mapping $\pi : V_C \rightarrow S$, recall we defined $W(\pi)  = \sum_{j \in V_C} d(\sigma_O(j), \pi(j))$ and, from now on, we let $\pi$ be a feasible assignment minimizing $W(\pi)$.
If $|O| < |S|$, we have to add one more condition to the definition of $\pi$. First, let $\mu : O \rightarrow S$ be a minimum-distance injection, we will show one exists for which $\mu(i) = i$ for each $i \in O \cap S$. In the case $|O| < |S|$, among all optimal choices for $\pi$
we further choose one that minimizes $|\{j \in V_C : \pi(j) \notin \mu(O)\}|$. If $|O| \geq |S|$, no extra conditions beyond minimizing $W(\pi)$ are needed for $\pi$.

Let $P = \pi(V_C)$, i.e. the image of $\pi$. We bound $F(S)$ by separately bounding $F(P)$ and $F(S-P)$. The bound on $F(S-P)$ works for both \flmi with uniform opening costs and for \flmi with group constraints,
however the analysis of the two cases differ significantly in how they bound $F(P)$.

\subsection{Supermodularity of $A(\cdot)$ and Bounding Connection Costs}
The purpose of this section is the following.
\begin{lemma}
\label{lem:connection-bound}
$C\le C^*+ F^*$
\end{lemma}

This bound would hold even in the more general case \flmi with non-uniform opening costs. Note, this is essentially the same sort of bound used in nearly all local search algorithms for facility location. In our case,
as with Capacitated Facility Location, it follows from a notion of supermodularity of $A(\cdot)$, where we recall $A(X)$ is the optimum connection cost of a feasible assignment to $X \subseteq V_C$.

That is, we prove the following:
\begin{lemma}\label{lem:super}
For all $X,Y \subseteq V_F$ we have $A(X) + A(Y) \le A(X \cup Y) + A(X \cap Y)$.
\end{lemma}
In particular, this implies whenever $X$ is feasible, $Y \supseteq X$, and $e \notin Y$ that we have
\begin{equation}
\label{eq:general-matroid-diminishing-improvement}
    A(Y)-A(Y\cup\{e\})
    \le
    A(X)-A(X\cup\{e\}).
\end{equation}
We also note the proof of Lemma \ref{lem:super} we present would also work to prove the generalization to where each $i \in V_F$ had its own matroid $\mathcal M_i$ and $A(X)$ was the optimal way to map $V_C$ to $i \in X$
ensuring each facility in $X$ receives an independent set in $\mathcal M_i$. But this generality is not needed for our results.

For now, we show how this is used to bound the opening costs in a standard way.
\begin{proof}[Proof of Lemma \ref{lem:connection-bound}]
For every \(o\in O\setminus S\), add-move local optimality gives
    $C-A(S\cup\{o\})\le F(\{o\}).$
Summing these inequalities over all \(o\in O\setminus S\), and using
Lemma~\ref{lem:super} in a standard way, we obtain
\begin{align*}
    C-A(S\cup O)
    &\le
    \sum_{o\in O\setminus S}
    \left(C-A(S\cup\{o\})\right)\\
    &\le
    F(O\setminus S).
\end{align*}

Here, the first bound is justified by adding $o \in O-S$ to $S$ one at a time. That is,
write \(O\setminus S=\{i_1,\ldots,i_m\}\) and let
\(S_t:=S\cup\{i_1,\ldots,i_t\}\) for $0 \leq t \leq m$.  Applying
\eqref{eq:general-matroid-diminishing-improvement} with
\(X=S\), \(Y=S_{t-1}\), and \(e=i_t\) gives
$ A(S_{t-1})-A(S_t)
    \le
    A(S)-A(S\cup\{i_t\}).$
Summing over \(t=1,\ldots,m\) and telescoping establishes the first inequality.

The optimal assignment \(\sigma_O\) remains feasible on the larger facility
set \(S\cup O\), so
    $A(S\cup O)\le C^*.$
Substituting this into the inequality above gives
$ C-C^*
    \le
    F(O\setminus S)
    \le
    F^*,$
as required.
\end{proof}
We remark this bound only used the add moves of the local search procedure.

~

\noindent
{\bf Supermodularity}\\
The rest of this section is dedicated to proving Lemma \ref{lem:super}. We assume $X \cap Y$ is feasible, otherwise the right-hand side of the supermodularity bound is $\infty$ and there is nothing else to show. Note, this implies $X,Y$ and $X \cup Y$ are also feasible.

Our strategy is the following:
\begin{itemize}
\item Use duality for matroid intersection to give a dual characterization of the optimal assignment cost $A(X)$, see Lemma \ref{lem:general-matroid-dual-representation} immediately below.
\item Use properties of the greedy algorithm for matroids to characterize how the term that maximizes over independent sets can be viewed as a layering of the costs, i.e. Lemma \ref{lem:layer}.
\item Finally, we point out how the corresponding integrals for the calculations of $A(X), A(Y), A(X \cup Y)$ and $A(X \cap Y)$ satisfy an appropriate supermodularity condition (Equation \eqref{eqn:meetjoin}) which directly translates to proving $A(\cdot)$ itself is supermodular.
\end{itemize}

To that end, recall that \(\mathcal M=(V_C,\mathcal I)\) is the matroid on the client set. Let \(r\) denote the rank function of $\mathcal M$. A feasible assignment requires the set of clients assigned to each open facility to be independent in \(\mathcal M\). We begin with a dual representation of \(A(X)\) for an arbitrary feasible facility set \(X\).
\begin{lemma}
\label{lem:general-matroid-dual-representation}
For every feasible facility set \(X\),
\begin{equation}
\label{eq:general-matroid-dual-representation}
    A(X)=
    \max_{p\in\mathbb R^{V_C}}
    \left\{
        \sum_{j\in V_C}p_j
        -\sum_{i\in X}
        \max_{I\in\mathcal I}
        \sum_{j\in I}\left(p_j-d(j,i)\right)
    \right\}.
\end{equation}
\end{lemma}
Many similar min-max relationships for weighted matroid intersection and union problems can be found in literature, e.g. \cite{schrijver2003}. We were unable to find a match for our exact setting, so we provide a proof here.
\begin{proof}
Using the rank function \(r\) of \(\mathcal M\), the LP formulation for
the assignment problem on a fixed facility set \(X\) is as follows:
\begin{equation}
\label{eq:general-matroid-rank-assignment-lp}
\begin{aligned}
    \min\quad&
        \sum_{i\in X}\sum_{j\in V_C}d(j,i)x_{ij}\\
    \text{s.t.}\quad&
        \sum_{i\in X}x_{ij}=1
        &&(j\in V_C),\\
    &\sum_{j\in U}x_{ij}\le r(U)
        &&(i\in X,\ U\subseteq V_C),\\
    &x_{ij}\ge0
        &&(i\in X,\ j\in V_C).
\end{aligned}
\end{equation}
The constraints define a face of the intersection of the client
partition matroid and the direct sum of one copy of \(\mathcal M\) for
each facility in \(X\). By Edmonds' matroid-intersection polytope
theorem~\cite{schrijver2003}, this formulation is integral. Since its integral feasible solutions are exactly
the feasible assignments to \(X\), its optimum is \(A(X)\).

Let \(p_j\in\mathbb R\) be the dual variable for the equality corresponding
to client \(j\), and let \(y_{i,U}\ge0\) be the dual variable for the rank
constraint indexed by \((i,U)\).  The dual of
\eqref{eq:general-matroid-rank-assignment-lp} is
\begin{equation}
\label{eq:general-matroid-rank-assignment-dual}
\begin{aligned}
    \max\quad&
        \sum_{j\in V_C}p_j
        -\sum_{i\in X}\sum_{U\subseteq V_C}r(U)y_{i,U}\\
    \text{s.t.}\quad&
        p_j-\sum_{U\ni j}y_{i,U}\le d(j,i)
        &&(i\in X,\ j\in V_C),\\
    &y_{i,U}\ge0
        &&(i\in X,\ U\subseteq V_C).
\end{aligned}
\end{equation}
Fix \(p\).  For each facility \(i\), maximizing the dual objective requires
minimizing
    $\sum_{U\subseteq V_C}r(U)y_{i,U}$
subject to
    $\sum_{U\ni j}y_{i,U}\ge p_j-d(j,i)
    ~(j\in V_C),
    y_{i,U}\ge0.$
The dual of this minimization problem is
\[
\begin{aligned}
\max \quad
& \sum_{j\in V_C}\left(p_j-d(j,i)\right)z_j \\[2mm]
\text{s.t.}\quad
& \sum_{j\in U} z_j \le r(U)
&& \forall\, U\subseteq V_C,\\
& z_j\ge 0
&& \forall\, j\in V_C.
\end{aligned}
\]
By the matroid independent-set polytope theorem, the rank-inequality
system is integral, and its integral points are exactly the characteristic
vectors of independent sets of \(\mathcal M\). Hence the maximum is
attained by the incidence vector of an independent set and equals
   $ \max_{I\in\mathcal I}
    \sum_{j\in I}\left(p_j-d(j,i)\right).$
Replacing the \(y\)-variables from
\eqref{eq:general-matroid-rank-assignment-dual} therefore yields
\eqref{eq:general-matroid-dual-representation}.
\end{proof}

We next express the facility contribution directly through the rank
function.  For any vector \(z\in\mathbb R^{V_C}\), the
layer-cake representation\footnote{The term \emph{layer-cake representation}
is standard in mathematical analysis and refers to
expressing a nonnegative function as an integral
of indicators of its superlevel sets.
See \cite{lieb2001analysis} section 1.13.},
together with the matroid greedy theorem, yields the following lemma
\begin{lemma}[Theorem 40.2 in \cite{schrijver2003}]
\label{lem:layer}
\begin{equation}
\label{eq:general-matroid-layer-cake}
    \max_{I\in\mathcal I}\sum_{j\in I}z_j
    =
    \int_0^\infty
    r\left(\{j\in V_C:z_j\ge t\}\right)\,dt.
\end{equation}
\end{lemma}
Note that \cite{schrijver2003} states it in terms of discrete sums weighted by gaps between different $z_j$-values, our statement is equivalent.

Applying \eqref{eq:general-matroid-layer-cake} to
\(z_j=p_j-d(j,i)\) gives
\begin{equation}
\label{eq:general-matroid-rank-layer-cake}
    \max_{I\in\mathcal I}
    \sum_{j\in I}\left(p_j-d(j,i)\right)
    =
    \int_0^\infty
    r\left(\{j\in V_C:p_j-d(j,i)\ge t\}\right)\,dt.
\end{equation}
In particular, the quantity on the left is coordinatewise nondecreasing in
\(p\).  Indeed, if \(a\le b\) coordinatewise, then for every \(t\ge0\),
\[
    \{j:a_j-d(j,i)\ge t\}
    \subseteq
    \{j:b_j-d(j,i)\ge t\},
\]
and rank monotonicity in \eqref{eq:general-matroid-rank-layer-cake} gives
\[
    \max_{I\in\mathcal I}
    \sum_{j\in I}\left(a_j-d(j,i)\right)
    \le
    \max_{I\in\mathcal I}
    \sum_{j\in I}\left(b_j-d(j,i)\right).
\]

We can now combine the dual representation with the layer-cake identity and the supermodularity of the matroid rank function to prove supermodularity of the connection-cost function.

\begin{proof}[Proof of Lemma \ref{lem:super}]
Let \(p\) and \(q\) be optimal dual vectors in
\eqref{eq:general-matroid-dual-representation} for \(X\) and \(Y\),
respectively. For each \(j\in V_C\), define
$    u_j:=\max\{p_j,q_j\},
    \ell_j:=\min\{p_j,q_j\}.$
 For a facility set
\(Z\) and a vector \(a\in\mathbb R^{V_C}\), let
\[
    L_Z(a):=
    \sum_{j\in V_C}a_j
    -\sum_{k\in Z}
    \max_{I\in\mathcal I}
    \sum_{j\in I}\left(a_j-d(j,k)\right).
\]
Thus, by \eqref{eq:general-matroid-dual-representation},
\(A(X)=L_X(p)\) and \(A(Y)=L_Y(q)\).

We first compare the two facility terms associated with a common facility
\(k\in X\cap Y\).  For every threshold \(t\ge0\), define
$    P_t^k:=\{j\in V_C:p_j-d(j,k)\ge t\},
    Q_t^k:=\{j\in V_C:q_j-d(j,k)\ge t\}.$
Because \(u_j=\max\{p_j,q_j\}\) and
\(\ell_j=\min\{p_j,q_j\}\),
\[
\begin{aligned}
    \{j:u_j-d(j,k)\ge t\}
        =P_t^k\cup Q_t^k,\quad
    \{j:\ell_j-d(j,k)\ge t\}
        =P_t^k\cap Q_t^k.
\end{aligned}
\]
For each fixed \(t\), supermodularity of the matroid rank function gives
$r(P_t^k)+r(Q_t^k)
    \ge
    r(P_t^k\cup Q_t^k)+r(P_t^k\cap Q_t^k).
$
Therefore, integrating
both sides preserves the inequality:

\begin{equation}\label{eqn:meetjoin}
    \int_0^\infty
        \left(r(P_t^k)+r(Q_t^k)\right)\,dt \ge
    \int_0^\infty
        \left(
            r(P_t^k\cup Q_t^k)+r(P_t^k\cap Q_t^k)
        \right)\,dt.
\end{equation}

Using \eqref{eq:general-matroid-rank-layer-cake} and the two threshold-set
identities above, for every \(k\in X\cap Y\) we obtain
\begin{equation}
\label{eq:general-matroid-common-facility}
    \max_{I\in\mathcal I}
        \sum_{j\in I}\left(p_j-d(j,k)\right)
    +\max_{I\in\mathcal I}
        \sum_{j\in I}\left(q_j-d(j,k)\right)
    \ge
    \max_{I\in\mathcal I}
        \sum_{j\in I}\left(u_j-d(j,k)\right)
    +\max_{I\in\mathcal I}
        \sum_{j\in I}\left(\ell_j-d(j,k)\right).
\end{equation}

For every \(k\in X\setminus Y\), we have \(\ell\le p\) coordinatewise, so
the coordinatewise monotonicity established above gives
  $  \max_{I\in\mathcal I}
    \sum_{j\in I}\left(p_j-d(j,k)\right)
    \ge
    \max_{I\in\mathcal I}
    \sum_{j\in I}\left(\ell_j-d(j,k)\right).$
Similarly, for every \(k\in Y\setminus X\), since \(\ell\le q\),
    $ \max_{I\in\mathcal I}
   \sum_{j\in I}\left(q_j-d(j,k)\right)
    \ge
    \max_{I\in\mathcal I}
    \sum_{j\in I}\left(\ell_j-d(j,k)\right).$

Combining these inequalities with
\eqref{eq:general-matroid-common-facility} and summing over all facilities
gives
\[
\begin{aligned}
    &\sum_{k\in X}
        \max_{I\in\mathcal I}
        \sum_{j\in I}\left(p_j-d(j,k)\right)
    +\sum_{k\in Y}
        \max_{I\in\mathcal I}
        \sum_{j\in I}\left(q_j-d(j,k)\right)\\
    &\ge
    \sum_{k\in X\cap Y}
    \left(
        \max_{I\in\mathcal I}
        \sum_{j\in I}\left(u_j-d(j,k)\right)
        +
        \max_{I\in\mathcal I}
        \sum_{j\in I}\left(\ell_j-d(j,k)\right)
    \right)\\
    &\qquad\quad\qquad\quad\qquad\quad\qquad\quad
    +\sum_{k\in X\setminus Y}
        \max_{I\in\mathcal I}
        \sum_{j\in I}\left(\ell_j-d(j,k)\right)
    +\sum_{k\in Y\setminus X}
        \max_{I\in\mathcal I}
        \sum_{j\in I}\left(\ell_j-d(j,k)\right)\\
    &=
    \sum_{k\in X\cup Y}
        \max_{I\in\mathcal I}
        \sum_{j\in I}\left(\ell_j-d(j,k)\right)
    +\sum_{k\in X\cap Y}
        \max_{I\in\mathcal I}
        \sum_{j\in I}\left(u_j-d(j,k)\right).
\end{aligned}
\]
Also, \(p_j+q_j=u_j+\ell_j\) for every client \(j\).  Therefore,
\[
\begin{aligned}
    A(X)+A(Y)
    &=L_X(p)+L_Y(q)\\
    &=
    \sum_{j\in V_C}(p_j+q_j)
    -\sum_{k\in X}
        \max_{I\in\mathcal I}
        \sum_{j\in I}\left(p_j-d(j,k)\right)
    -\sum_{k\in Y}
        \max_{I\in\mathcal I}
        \sum_{j\in I}\left(q_j-d(j,k)\right)\\
    &\le
    \sum_{j\in V_C}(\ell_j+u_j)
    -\sum_{k\in X\cup Y}
        \max_{I\in\mathcal I}
        \sum_{j\in I}\left(\ell_j-d(j,k)\right)
    -\sum_{k\in X\cap Y}
        \max_{I\in\mathcal I}
        \sum_{j\in I}\left(u_j-d(j,k)\right)\\
    &=L_{X\cup Y}(\ell)+L_{X\cap Y}(u)\\
    &\le A(X\cup Y)+A(X\cap Y).
\end{aligned}
\]

%
%


Since the dual variables in
\eqref{eq:general-matroid-dual-representation} are unrestricted,
\(\ell=\min\{p,q\}\) and \(u=\max\{p,q\}\) are also feasible dual vectors. Hence,
    $L_{X\cup Y}(\ell)\le A(X\cup Y),$
    and
    $L_{X\cap Y}(u)\le A(X\cap Y).$

\end{proof}

\subsection{Bounding $F(S\setminus P)$}
The preceding section showed $C \leq C^* + F^*$. To complete analysis of the local search algorithms, we need to bound $F$.
The primary difficultly is bounding $F(P)$, we provide that bound in future sections dedicated to the particular cases of \flmi with uniform opening costs and \flmi with group constraints.
Bounding $F(S\setminus P)$ is much more straightforward and can be done in a way that is common to all \flmi problems.

We begin with the following observation.
\begin{lemma}\label{lem:AP-bound}
    \(A(P)\le C+2C^*.\)
\end{lemma}
\begin{proof}
Since $\sigma_S$ is a feasible assignment to $S$ and $\pi$ minimizes
$W$, the triangle inequality gives
\[
    W(\pi)
    \le \sum_{j\in V_C}d\left(\sigma_O(j),\sigma_S(j)\right)
    \le C^*+C.
\]
Then, since $\pi$ is a feasible assignment supported on $P=\pi(V_C)$,
\[
\begin{aligned}
    A(P)
    &\le \sum_{j\in V_C}d\left(j,\pi(j)\right)\\
    &\le \sum_{j\in V_C}
       d\left(j,\sigma_O(j)\right)
            +d\left(\sigma_O(j),\pi(j)\right)\\
    &=C^*+W(\pi)\\
    &\le C+2C^*.
\end{aligned}
\]
\end{proof}

From this, we complete our analysis.
\begin{lemma}
\label{lem:NOPfacility-bound}
Let \(P=\pi(V_C)\). If \(S\) is locally optimal with respect to delete
moves, then
\[
    F(S\setminus P)\le 2C^*.
\]
\end{lemma}

\begin{proof}
Recall \(P=\pi(V_C)\). For every \(s\in S\setminus P\), the set
\(S\setminus\{s\}\) is feasible, since it still contains the feasible
facility set \(P\). Hence, delete-move local optimality gives
    $F(\{s\})
    \le
    A(S\setminus\{s\})-A(S).$
Summing over all \(s\in S\setminus P\), and applying supermoduarity of $A(\cdot)$ in a standard way we obtain
\begin{align*}
    F(S\setminus P)
    &=
    \sum_{s\in S\setminus P}F(\{s\})\\
    &\le
    \sum_{s\in S\setminus P}
    \left(A(S\setminus\{s\})-A(S)\right)\\
    &\le
    A(P)-A(S).
\end{align*}
The first inequality is because $S$ is a locally optimum solution, so deleting facilities does not result in a cheaper solution.
To justify the second inequality, write
\(S\setminus P=\{s_1,\ldots,s_m\}\) and put
   $ Z_t:=S\setminus\{s_1,\ldots,s_t\}
    ,(t=0,1,\ldots,m).$
Every \(Z_t\) is feasible because it contains the feasible set \(P\).  Since
\(Z_t\subseteq S\setminus\{s_t\}\), applying
\eqref{eq:general-matroid-diminishing-improvement} with
$   X=Z_t,
    Y=S\setminus\{s_t\},
    e=s_t$
gives
   $ A(S\setminus\{s_t\})-A(S)
    \le
    A(Z_t)-A(Z_{t-1}).$
Summing over \(t=1,\ldots,m\) and telescoping from \(Z_0=S\) to \(Z_m=P\) established the bound.

Since \(A(P)\le C+2C^*\) and \(A(S)=C\), it follows that
    $F(S\setminus P)\le 2C^*,$
which proves the claim.
\end{proof}
We remark this bound only used the delete moves in the local search procedure.


\subsection{Properties of $P$}
To help bound $F(P)$, there are further properties useful for the proofs of both Theorems \ref{thm:general-matroid-approximation} and \ref{thm:partition-matroid-nonuniform}.
It will be helpful to recall the definition of $P$ and the manner in which we constructed $\pi$ and $\eta$ from the start of Section \ref{sec:notation}, we rely heavily on their descriptions here.
The main result of this section is to show the following.

\begin{theorem}\label{thm:pbound}
If $|O| < |S|$, then $P \subseteq \mu(O)$.
\end{theorem}

In Section \ref{sec:notation}, we mentioned the minimum-cost injection $\mu : O \rightarrow S$ can be selected so that $\mu(i) = i$ for each $i \in O \cap S$.
We now prove it.

\begin{lemma}\label{lem:common-fixed-injection}
Suppose \(|O|<|S|\). There exists a minimum-cost injection \(\mu:O\to S\) such that
  $  \mu(i)=i$ for every $ q\in S\cap O$.
\end{lemma}
\begin{proof}
Let \(\mu:O\to S\) be a minimum-cost injection, chosen to maximize the
number of facilities \(i\in S\cap O\) satisfying \(\mu(i)=i\).

If $\mu(i) = s \neq i$ for some $i \in S \cap O$, we can update $\mu$ as follows: if $i \notin \mu(O)$ we can change $\mu(i)$ from $s$ to $i$, which does not increase
the cost as $d(i,i) = 0 \leq d(i,s)$. Otherwise, if $\mu(o) = i$ for some $o \in O-\{i\}$ we can swap $\mu(i)$ and $\mu(o)$ so that $\mu(i) = i$ and $\mu(o) = s$.
The cost changes by $d(i,i) + d(o,s) - d(i,s) - d(o,i) \leq 0$ by the triangle inequality.

In either case, we have another minimum-cost mapping $\mu : O \rightarrow S$ with more fixed
points than before, contradicting our initial choice of $\mu$. So $\mu(i) = i$ already holds for each $i \in S \cap O$.
\end{proof}

We also require the following supporting result.
\begin{lemma}\label{fact:injection}
Let \(\mu:O\to S\) be an injection minimizing
    $\sum_{o\in O} d(o,\mu(o)).$
For every unmatched facility \(t\in S\setminus\mu(O)\) and every sequence
\(o_0,o_1,\ldots,o_\ell\in O\), with repetitions allowed,
\begin{equation}\label{eq:alternating-walk}
    d(o_0,t)
    +\sum_{r=1}^{\ell} d\left(o_r,\mu(o_{r-1})\right)
    \ge
    \sum_{r=0}^{\ell} d\left(o_r,\mu(o_r)\right).
\end{equation}
\end{lemma}

\begin{proof}
First suppose that \(o_0,o_1,\ldots,o_\ell\) are distinct. Define a new
injection \(\mu':O\to S\) by
\[
    \mu'(o_0):=t,
    \qquad
    \mu'(o_r):=\mu(o_{r-1})
    \quad(r=1,\ldots,\ell),
\]
and let \(\mu'(o)=\mu(o)\) for all other \(o\in O\).
Since \(t\notin\mu(O)\) and the \(o_r\)'s are distinct, \(\mu'\) is an
injection. By the minimality of \(\mu\),
$    \sum_{r=0}^{\ell}d\left(o_r,\mu(o_r)\right)
    \le
    d(o_0,t)
    +\sum_{r=1}^{\ell}
        d\left(o_r,\mu(o_{r-1})\right),$
which is exactly \eqref{eq:alternating-walk}.

Now consider a sequence with repetitions. Choose \(i<j\) with
\(o_i=o_j\) such that
$    o_i,o_{i+1},\ldots,o_{j-1}$
are distinct. Cyclically shift their images under \(\mu\) by setting
\[
    \mu'(o_i):=\mu(o_{j-1}),
    \qquad
    \mu'(o_r):=\mu(o_{r-1})
    \quad(i<r<j),
\]
and leaving all other images unchanged. This is again an injection, so
the minimality of \(\mu\) gives
\[
    \sum_{r=i+1}^{j}
        d\left(o_r,\mu(o_{r-1})\right)
    \ge
    \sum_{r=i+1}^{j}
        d\left(o_r,\mu(o_r)\right),
\]
where \(o_j=o_i\).

Thus the closed segment
   $ o_i,o_{i+1},\ldots,o_j$
contributes a nonnegative amount to the difference between the two sides
of \eqref{eq:alternating-walk}. We may therefore remove this closed
segment from the sequence. Repeating this operation eliminates all
repetitions, after which the first part of the proof applies. Hence
\eqref{eq:alternating-walk} holds for every sequence, including those
with repetitions.
\end{proof}

Finally, we prove the main result of this section.
\begin{lemma}
\label{lem:matroid-image-subset}
If $|O| < |S|$, then $P \subseteq \mu(O)$.
\end{lemma}

\begin{proof}
Suppose, for contradiction, that
$P\not\subseteq\mu(O)$.
Then there exists a client $j_0\in\VC$ such that
$t:=\pi(j_0)\notin\mu(O)$.

We construct a chain of reassignments starting from $j_0$,
using an exchange argument similar to those in combinatorial matroid
intersection algorithms.
As illustrated in Figure~\ref{fig:reassignment-chain},
the idea is to move each client $j_r$ to its target
facility $p_r=\mu(\sigma_O(j_r))$.
If inserting a client violates independence, we use a
fundamental circuit to displace another client and
continue the chain.
Unlike a general search for an exchange sequence,
the target facility of each displaced client is
prescribed, although we may choose which client
to displace.
The process terminates when a client can be inserted
without displacing another client.
\begin{figure}[h]
    \centering
    \includegraphics[width=\linewidth]{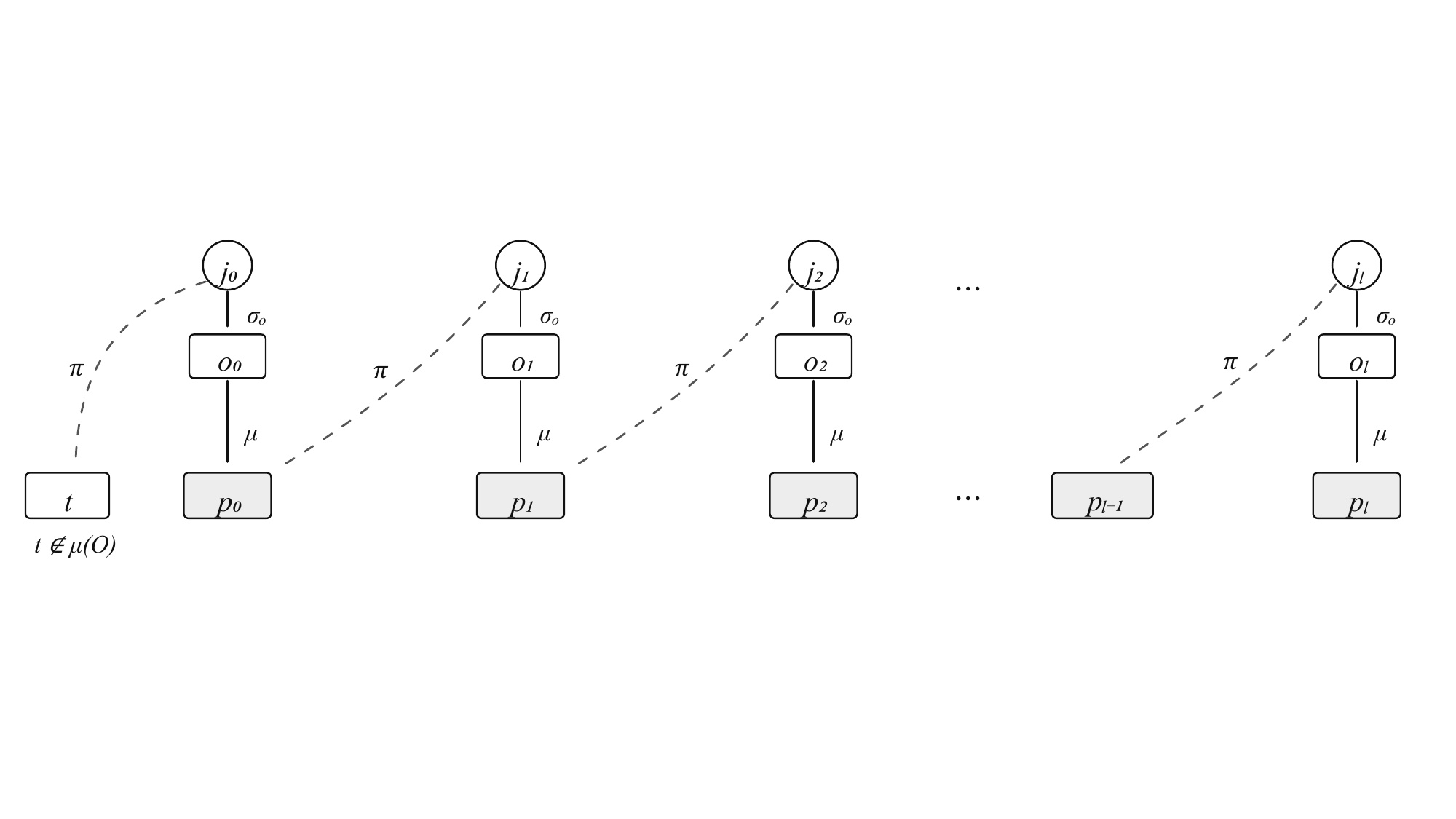}
    \caption{Illustration of the reassignment chain.
    For each client $j_r$, let $o_r=\sigma_O(j_r)$
    and $p_r=\mu(o_r)$.
    Solid edges represent the mappings $\sigma_O$ and $\mu$,
    while dashed edges represent the assignment
    $\pi$, where $\pi(j_0)=t\notin\mu(O)$ and
    $\pi(j_r)=p_{r-1}$ for $r\geq 1$.
    The facilities $p_r$ need not be distinct.}
    \label{fig:reassignment-chain}
\end{figure}

For each $p\in\mu(O)$, define
\[
    J_p:=\{j\in\VC:\mu(\sigma_O(j))=p\}.
\]
Thus $J_p$ is the set of clients whose target is $p$.
Since $\mu$ is injective, if $p=\mu(o)$, then
$J_p=\sigma_O^{-1}(o)$, which is independent in $M$
by the feasibility of $\sigma_O$.

We begin by removing $j_0$ from $t$.
At step $r$, let $j_r$ be the client currently being
reassigned, and write
\[
    o_r:=\sigma_O(j_r), \qquad p_r:=\mu(o_r).
\]
We maintain the following invariants at the beginning
of each step:
\begin{enumerate}
    \item[(i)] $j_r$ is the only unassigned client, and
    the clients currently assigned to each facility
    form an independent set in $M$.

    \item[(ii)] Every previously processed client
    $j_a$, $0\leq a<r$, is assigned to its target
    facility $p_a$. All clients outside
    $\{j_0,\ldots,j_r\}$ retain their original
    assignments under $\pi$.
\end{enumerate}
Both invariants hold initially for $r=0$.

Let $I_{p_r}$ be the current set of clients assigned
to $p_r$. If $I_{p_r}\cup\{j_r\}$ is independent,
we assign $j_r$ to $p_r$ and terminate.

Otherwise, since $I_{p_r}$ is independent,
$I_{p_r}\cup\{j_r\}$ contains a unique fundamental
circuit $C_r$. Since $j_r\in J_{p_r}$ and
$J_{p_r}$ is independent, $C_r$ cannot be contained
in $J_{p_r}$. Hence we may choose
\[
    j_{r+1}\in C_r\setminus J_{p_r}.
\]
In particular, $j_{r+1}\neq j_r$, so
$j_{r+1}\in I_{p_r}$. By the fundamental-circuit
property,
\[
    (I_{p_r}\setminus\{j_{r+1}\})\cup\{j_r\}
\]
is independent.

Moreover, $j_{r+1}$ cannot be one of the previously
processed clients, since those clients are already
at their target facilities and hence belong to
the corresponding sets $J_p$.
Thus $j_{r+1}$ has not been moved before, and
\[
    \pi(j_{r+1})=p_r=\mu(o_r).
\]
We replace $j_{r+1}$ by $j_r$ at $p_r$ and continue
with $j_{r+1}$. Both invariants are preserved.

This process must terminate. At each step, $j_r$
is placed at its target facility, while any displaced
client $j_{r+1}$ lies outside $J_{p_r}$ and therefore
was not at its target. Furthermore, a client already
at its target can never be displaced, since each
displaced client is chosen outside the corresponding
set $J_p$. Hence the number of clients assigned to
their target facilities strictly increases at each step.
Since there are finitely many clients, the process
terminates after finitely many steps.

Let $j_0,j_1,\ldots,j_\ell$ be the resulting chain,
where $j_\ell$ is inserted into $p_\ell$ without
displacing another client. By invariant (ii), these
clients are distinct and
\[
    \pi(j_r)=\mu(o_{r-1}),
    \qquad r=1,\ldots,\ell.
\]

Let $\pi'$ be the resulting assignment. Equivalently,
\[
    \pi'(j_r)=\mu(o_r),
    \qquad r=0,\ldots,\ell,
\]
while all other clients retain their assignments
under $\pi$.

By invariant (i) and the final insertion,
$\pi'$ is feasible. This remains true even if the
same facility is visited more than once, since
each exchange uses the fundamental circuit of
the current independent set at that facility.

The contribution of the affected clients to
$W(\pi)$ is
\[
    d(o_0,t)
    +\sum_{r=1}^{\ell}
        d(o_r,\mu(o_{r-1})),
\]
whereas their contribution to $W(\pi')$ is
\[
    \sum_{r=0}^{\ell}d(o_r,\mu(o_r)).
\]
By Lemma~\ref{fact:injection},
\[
    W(\pi')\leq W(\pi).
\]

Finally, $j_0$ moves from $t\notin\mu(O)$ to a
facility in $\mu(O)$, while every other client
in the chain starts and ends at a facility in
$\mu(O)$. Hence
\[
    q(\pi')=q(\pi)-1.
\]
If $W(\pi')<W(\pi)$, this contradicts the
minimality of $W(\pi)$. If equality holds,
it contradicts the tie-breaking minimality
of $q(\pi)$.

Therefore, $q(\pi)=0$, and hence
$P\subseteq\mu(O)$.
\end{proof}

\section{\flmi With Uniform Opening Costs}
\label{sec:gen-matroid-unif-cost}

Given $C \leq C^* + F^*$ (Lemma~\ref{lem:connection-bound}) and $F(S\setminus P) \leq 2C^*$ (Lemma~\ref{lem:AP-bound}),
we only have to bound $F(P)$. But this is quite straightforward in this case.
\begin{lemma}
For \flmi with uniform opening costs, a local minimum solution satisfies $F(P) \leq F^*$.
\end{lemma}
\begin{proof}
We claim $|P| \leq |O|$, from which it follows $F(P) = f \cdot |P| \leq f \cdot |O| = F^*$.
If $|S| \leq |O|$ this is immediate because $P \subseteq S$.
On the other hand, by Lemma \ref{lem:matroid-image-subset}
if $|S| > |O|$ then $P \subseteq \mu(O)$. Since $\mu$ is a one-to-one mapping, this also means $|P| \leq |O|$.
In either case, we are finished.
\end{proof}

Overall, we have that $C \leq C^* + F^*$ and $F \leq 2C^* + F^*$, which combines to show $\cost(S) = C + F \leq 3C^* + 2F^* \leq 3 \cdot OPT$.
The standard rebalancing trick works here to improve the ratio.
The local search analysis does not actually require optimality of the solution $O$, it shows for any feasible solutions $S, S' \subseteq V_F$ where $S$ is a local minimum
that $A(S) \leq A(S') + F(S')$ and $F(S) \leq 2A(S') + F(S')$.

Let $\lambda > 0$ be a value to be decided soon and run local search except with uniform facility costs $f/\lambda$. Let $C, F/\lambda$ denote the connection and facility costs of the local optimum solution $S$
and $C^*, F^*/\lambda$ the connection and facility costs of the global optimum solution $O$ (to the instance before scaling). Then we have $C \leq C^* + F^*/\lambda$ and $F/\lambda \leq 2C^* + F^*/\lambda$
showing
\[ C + F \leq (1 + 2\lambda)C^* + (1 + 1/\lambda)F^* \]
Setting $\lambda = 1/\sqrt{2}$ shows $\cost(S) = C + F \leq (1+\sqrt 2)(C^* + F^* = (1 + \sqrt 2) OPT$.

This proves Theorem \ref{thm:general-matroid-approximation}.
Note, to ensure the algorithm runs using rational values, we can set $\lambda$ to be a fraction arbitrarily close to $1/\sqrt{2}$. This, and the adaptation of local search to ensure it uses a polynomial number of steps are
the cause of the introduction of $\epsilon$ in the approximation guarantee.

%
%
%
%
%
%

\section{\flmi with Group Constraints and Non-uniform Opening Costs}
\label{sec:group-non-uniform}

In this section, we consider \flmi with group constraints
and non-uniform facility-opening costs.
Recall that $\mathcal M$ is a partition matroid over $V_C$.
We call its parts \emph{groups} and denote the clients
in group $g$ by $G_g$. The matroid constraint requires
that each facility serves at most $b_g$ clients from $G_g$.

As in the uniform-opening-cost case, the results from
Section~\ref{sec:notation} allow us to bound
$F(S\setminus P)$ by $2C^*$.
The main difficulty is bounding $F(P)$.
Unlike the uniform-cost case, the bound $|P|\leq |O|$
is no longer sufficient to control the total opening cost.

We therefore construct a bounded-congestion mapping
$\nu:P\to O$ and consider test swaps that replace
facilities in $P$ with their images under $\nu$.
The mapping will be chosen so that each non-common
facility in $O$ has at most eight preimages.
This allows us to control the total opening-cost
contribution when summing the inequalities associated
with these swaps. The more challenging part is
controlling the cost of reassigning clients.

We first construct the mapping independently of the
group capacities. To analyze the test swaps, we begin
with the simpler case where $b_g=1$ for every group.
In this case, each facility serves at most one client
from each group, allowing us to bound the reassignment
costs separately for each group and construct the
injections needed for our analysis.
We then extend the result to arbitrary group capacities
in Subsection~\ref{subsec:partition-extension}
by refining each group into capacity-one groups.

\subsection{Bounded-congestion mapping}
\label{subsec:Bounded-congestion mapping}

Recall from Section~\ref{sec:notation} that we chose
a feasible assignment $\pi:V_C\to S$ minimizing $W(\pi)$.
When $|O|<|S|$, we first choose a minimum-cost injection
$\mu:O\to S$ as in
Lemma~\ref{lem:common-fixed-injection},
which satisfies $\mu(q)=q$ for every $q\in S\cap O$.
We then choose $\pi$ among the $W$-minimizing assignments
as in Lemma~\ref{lem:matroid-image-subset}.

Write
\[
    P_0:=\pi(V_C).
\]
The set $P_0$ need not contain every facility in
$S\cap O$. To handle these common facilities,
we slightly enlarge it by defining
\[
    P:=P_0\cup(S\cap O).
\]
Since $P_0\subseteq S$, we have
\[
    Q:=P\cap O
      =(P_0\cap O)\cup(S\cap O)
      =S\cap O.
\]

We also have $|P|\leq |O|$.
If $|O|\geq |S|$, this follows from $P\subseteq S$.
Otherwise, Lemma~\ref{lem:matroid-image-subset}
gives $P_0\subseteq\mu(O)$.
Since $\mu(q)=q$ for every $q\in S\cap O$,
we also have $S\cap O\subseteq\mu(O)$.
Consequently,
\[
    P=P_0\cup(S\cap O)\subseteq\mu(O),
\]
and hence
\[
    |P|\leq|\mu(O)|=|O|.
\]

We now construct a mapping
\[
    \nu:P\to O.
\]
Unlike the injection used in the uniform-opening-cost
analysis, this mapping is not required to be injective
or surjective. Instead, each facility in $O\setminus Q$
is allowed to receive up to eight elements
\footnote{
The constants $3$ in the potential~\eqref{eq:potential}
and $8$ in the congestion bound are chosen to optimize
the approximation ratio obtained from our analysis.
More generally, one may replace them by parameters
$\alpha$ and $\beta$, respectively, and optimize the
resulting bound. We omit the straightforward calculation.
}
of $P\setminus Q$.

Fix $\nu(q)=q$ for every $q\in Q$.
These common facilities will not be used as targets
for other points. We refer to $O\setminus Q$ as
\emph{targets} and give each target eight available
\emph{slots}.

Repeatedly select a \emph{globally shortest} pair
$(s,o)$ among unassigned $s\in P\setminus Q$
and targets $o\in O\setminus Q$ with an available
slot. Set $\nu(s)=o$ and consume one slot.
Ties may be arbitrary. There are enough slots because
\[
    |P\setminus Q|\leq|O\setminus Q|.
\]

For any $T\subseteq P$, let $\mathcal E(T,O)$
denote the family of injections from $T$ to $O$.
Since $|T|\leq|P|\leq|O|$, this family is nonempty.
Define
\[
    \mc(T,O):=
    \min_{\eta\in\mathcal E(T,O)}
    \sum_{s\in T}d(s,\eta(s)).
\]

We emphasize that the greedy construction fixes
the map $\nu$ before any subset $T\subseteq P$
is selected. This is important because we will
use the same mapping for all group-specific subsets
in the test-swap analysis. The following is our
main technical bound on the reassignment costs.

\begin{lemma}[Simultaneous subset bound]
\label{lem:simultaneous-subset-bound}
The single map $\nu$ constructed above satisfies,
for every subset $T\subseteq P$,
\begin{equation}
\label{eq:subset}
    \sum_{s\in T}
    \min\{d(s,\nu(s)),\dist(s,P\setminus T)\}
    \leq 6\mc(T,O),
\end{equation}
where $\dist(s,\emptyset):=+\infty$.
Moreover, $\nu(q)=q$ for every $q\in Q$,
no other point maps to $Q$, and every
$o\in O\setminus Q$ has at most eight preimages.
\end{lemma}

\begin{proof}
The properties of $\nu$ on $Q$ and the congestion
bound follow directly from the construction.
We focus on proving~\eqref{eq:subset}.

Fix a nonempty $T$; the empty case is immediate.
For every $s\in T\cap Q$,
\[
    \min\{d(s,\nu(s)),\dist(s,P\setminus T)\}=0.
\]

Consider the following potential function:
for any point $x$,
\begin{equation}
\label{eq:potential}
    h(x)=
    \max\left\{
        0,
        \max_{v\in T}
        \left(
            \frac{
                \min\{d(v,\nu(v)),\dist(v,P\setminus T)\}
            }{3}
            -d(v,x)
        \right)
    \right\}.
\end{equation}

The function $h$ is nonnegative and 1-Lipschitz,
since it is the pointwise maximum of the zero
function and a family of 1-Lipschitz functions.
Moreover, for every $s\in T$,
\[
    h(s)\geq
    \frac{
        \min\{d(s,\nu(s)),\dist(s,P\setminus T)\}
    }{3}.
\]

For any injection $\eta:T\to O$,
\begin{equation}
\label{eqn:ho}
\begin{aligned}
    \sum_{s\in T}d(s,\eta(s))
    &\geq
    \sum_{s\in T}h(s)
    -\sum_{s\in T}h(\eta(s))\\
    &\geq
    \sum_{s\in T}h(s)
    -\sum_{o\in O}h(o).
\end{aligned}
\end{equation}

The first inequality follows from the
1-Lipschitz property
$h(s)-h(\eta(s))\leq d(s,\eta(s))$.
The second follows because $\eta$ is injective
and $h$ is nonnegative.
Thus, the last expression in~\eqref{eqn:ho}
is a lower bound on $\mc(T,O)$.

If $q\in Q\setminus T$, then
\[
    \min\{d(v,\nu(v)),\dist(v,P\setminus T)\}
    \leq \dist(v,P\setminus T)
    \leq d(v,q)
\]
for every $v\in T$, and hence $h(q)=0$.
The contributions from $Q\cap T$ cancel
in~\eqref{eqn:ho}. Consequently,
\begin{equation}
\label{eq:dual}
    \mc(T,O)\geq
    \sum_{s\in T\setminus Q}h(s)
    -\sum_{o\in O\setminus Q}h(o).
\end{equation}

We next bound the total potential of the targets. We only need to consider targets $o\in O\setminus Q$ with $h(o)>0$, since targets with zero potential do not contribute to the sum. Fix such a target $o$ and choose $v\in T$ attaining the positive maximum in~\eqref{eq:potential}. For such an $o$,
\[
    d(v,o)<
    \frac{
        \min\{d(v,\nu(v)),\dist(v,P\setminus T)\}
    }{3}
    \leq\frac{d(v,\nu(v))}{3}.
\]
In particular, $v\notin Q$.

The target $o$ must already have had all eight
slots filled when $v$ was assigned.
Otherwise, the pair $(v,o)$ would have been
eligible and strictly shorter than
$(v,\nu(v))$, contradicting the greedy choice.

Let $U_o$ be the eight points assigned to $o$.
They were all assigned before $v$, while $v$
was still unassigned. At each such assignment,
the pair $(v,o)$ was eligible.
Global shortest-pair selection therefore gives
\begin{equation}
\label{eq:small-edges}
    d(u,o)\leq d(v,o)
    \qquad(u\in U_o).
\end{equation}

In particular,
\[
\begin{aligned}
    d(v,u)
    &\leq d(v,o)+d(u,o)\\
    &\leq 2d(v,o)\\
    &<
    \min\{d(v,\nu(v)),\dist(v,P\setminus T)\}\\
    &\leq\dist(v,P\setminus T).
\end{aligned}
\]
Thus every $u\in U_o$ lies in $T$,
and is noncommon by construction.

Furthermore,
\[
\begin{aligned}
    \dist(u,P\setminus T)
    &\geq \dist(v,P\setminus T)-d(v,u)\\
    &\geq
    \min\{d(v,\nu(v)),\dist(v,P\setminus T)\}
    -2d(v,o)\\
    &>d(v,o)\\
    &\geq d(u,o)
    =d(u,\nu(u)).
\end{aligned}
\]
Therefore,
\[
    \min\{d(u,\nu(u)),\dist(u,P\setminus T)\}
    =d(u,\nu(u))
    =d(u,o).
\]
If $P\setminus T$ is empty, the same conclusion
follows directly from
$\dist(u,P\setminus T)=+\infty$.

The potential now supplies two inequalities:
\[
    h(u)\geq\frac{d(u,o)}{3},
    \qquad
    h(u)\geq h(o)-d(u,o).
\]
Adding the second inequality to three times
the first yields
\[
    4h(u)\geq h(o).
\]
There are eight points in $U_o$, so
\[
    \sum_{u\in U_o}h(u)\geq 2h(o).
\]

The sets $U_o$ are disjoint for distinct targets.
Since every $U_o$ in this argument lies in
$T\setminus Q$, we obtain
\begin{equation}
\label{eq:target-potential}
    \sum_{o\in O\setminus Q}h(o)
    \leq
    \frac12\sum_{s\in T\setminus Q}h(s).
\end{equation}

Substituting into~\eqref{eq:dual} gives
\[
\begin{aligned}
    \mc(T,O)
    &\geq
    \frac12\sum_{s\in T\setminus Q}h(s)\\
    &\geq
    \frac16\sum_{s\in T\setminus Q}
    \min\{d(s,\nu(s)),\dist(s,P\setminus T)\}.
\end{aligned}
\]
Since the summand is zero for $s\in T\cap Q$,
\[
    \mc(T,O)
    \geq
    \frac16\sum_{s\in T}
    \min\{d(s,\nu(s)),\dist(s,P\setminus T)\}.
\]
This proves~\eqref{eq:subset}.
\end{proof}

The map $\nu$ is fixed independently of $T$,
so the preceding inequality holds simultaneously
for all group-specific subsets $T_g$.
Moreover, the bound of eight preimages per
non-common target controls the total opening-cost
contribution across the test swaps.
We now use these properties to bound $F(P)$.

\subsection{Feasible test swaps and the opening-cost bound}
\label{subsec:capacity-one}

We first consider the special case where every
group has capacity one, i.e., $b_g=1$ for all $g$.
Thus, each facility serves at most one client
from each group. This allows us to analyze
the reassignment costs separately for each group.
We will extend the result to arbitrary group
capacities in Subsection~\ref{subsec:partition-extension}.

Recall that
\[
    Q=P\cap O=S\cap O.
\]
Hence, for every $s\in P\setminus Q$,
\[
    \nu(s)\in O\setminus Q=O\setminus S,
\]
so $\nu(s)$ is not open in the local solution $S$.

For each group $g$, define the local facilities
inside $P$ serving a client from $G_g$ by
\[
    T_g=
    \{s\in P:\sigma_S^{-1}(s)\cap G_g
    \neq\emptyset\}.
\]

For the test swap
$S'=S\setminus\{s\}\cup\{\nu(s)\}$,
keep all clients outside $\sigma_S^{-1}(s)$
unchanged.
For the unique client of a group $G_g$ at $s$,
choose whichever of the following has smaller
facility-to-facility distance:
\begin{enumerate}[label=(\roman*)]
    \item the newly opened facility $\nu(s)$;
    \item a nearest facility in $P\setminus T_g$,
    when this set is nonempty.
\end{enumerate}

The second facility is open and currently unused
by group $g$. Since there is at most one client
from each group at $s$, either reassignment
respects the capacity-one constraint.
Different groups may use the same backup facility
or the same newly opened facility.
Hence the reassignment is feasible and moves
only clients of the closed facility.

The connection-cost increase from this swap
is at most
\[
    w_s:=
    \sum_{g:s\in T_g}
    \min\{d(s,\nu(s)),\dist(s,P\setminus T_g)\}.
\]
The triangle inequality justifies this bound
separately for every moved client.
Local optimality implies
\begin{equation}
\label{eq:swap}
    F(\{s\})
    \leq F(\{\nu(s)\})+w_s
    \qquad(s\in P\setminus Q).
\end{equation}

For each group $g$, the clients locally served
at the facilities in $T_g$ induce an injection
from $T_g$ into $O$: map each $s\in T_g$
to the optimal facility of its unique
group-$g$ client.

This mapping is injective because $\sigma_O$
also satisfies the capacity-one constraint.
Therefore,
\[
    \mc(T_g,O)\leq
    \sum_{\substack{
        j\in G_g\\
        \sigma_S(j)\in P
    }}
    d(\sigma_S(j),\sigma_O(j)).
\]

The same mapping $\nu$ works for all groups,
so Lemma~\ref{lem:simultaneous-subset-bound}
and another triangle inequality give
\begin{align}
\label{eq:movement}
    \sum_{s\in P\setminus Q}w_s
    &\leq 6\sum_g\mc(T_g,O)\nonumber\\
    &\leq
    6\sum_{\sigma_S(j)\in P}
    d(\sigma_S(j),\sigma_O(j))\nonumber\\
    &\leq 6(C+C^*).
\end{align}
There is no separate opening-cost charge per group.

Summing~\eqref{eq:swap} and charging common
facilities directly yields
\begin{align}
\label{eq:inside}
    F(P)
    &\leq
    F(Q)+8F(O\setminus Q)+6(C+C^*)\nonumber\\
    &\leq 8F^*+6(C+C^*).
\end{align}

The preceding add- and delete-move bounds
continue to apply in the non-uniform-opening-cost
setting.
Since $P_0\subseteq P$,
Lemma~\ref{lem:AP-bound} gives
\[
    A(P)\leq A(P_0)\leq C+2C^*.
\]
Moreover, since $P$ contains $P_0$, it supports
the feasible assignment $\pi$.
Therefore, the proof of
Lemma~\ref{lem:NOPfacility-bound}
applies and gives
\[
    F(S\setminus P)
    \leq A(P)-A(S)
    \leq 2C^*.
\]
Lemma~\ref{lem:connection-bound} gives
\[
    C\leq C^*+F^*.
\]

Combining these bounds with~\eqref{eq:inside},
we obtain
\[
\begin{aligned}
    F
    &=F(S\setminus P)+F(P)\\
    &\leq 2C^*+8F^*+6(C+C^*)\\
    &=8(F^*+C^*)+6C.
\end{aligned}
\]

We can now complete the analysis for
capacity-one group constraints.

\begin{lemma}
\label{lem:capacity-one-nonuniform}
Suppose every group has capacity one.
Under arbitrary nonnegative facility-opening
costs, every solution $S$ that is locally
optimal with respect to add, delete, and
single-swap moves satisfies
\[
    \cost(S)\leq 15\cost(O),
\]
where $O$ is an optimal solution.
\end{lemma}

\begin{proof}
Using the bounds above,
\[
\begin{aligned}
    C+F
    &\leq 8(F^*+C^*)+7C\\
    &\leq 15(C^*+F^*),
\end{aligned}
\]
where the second inequality follows from
$C\leq C^*+F^*$.
\end{proof}

The capacity-one assumption is used in the
test-swap analysis above. It ensures that
each affected group contributes at most
one client to the reassignment cost at
a facility, and that the mapping from
$T_g$ to $O$ is injective.

For general capacities $b_g>1$, these
properties need not hold directly.
However, we can reduce the analysis to
the capacity-one case by refining each
group, as we show next.

\subsection{Extension to General Partition Matroids}
\label{subsec:partition-extension}

We now extend the preceding result to
general partition matroids, where each
group $G_g$ may have capacity $b_g>1$.

The idea is to refine each group into
smaller groups of capacity one.
We choose this refinement so that the
minimum-cost assignments of both the
local and optimal solutions remain feasible.
The analysis from Subsection~\ref{subsec:capacity-one}
can then be applied to the refined instance.


\begin{proof}[Proof of
Theorem~\ref{thm:partition-matroid-nonuniform}]

Let $S$ be a locally optimal solution and
$O$ an optimal solution.
Fix minimum-cost feasible assignments
$\sigma_S$ and $\sigma_O$ for $S$ and $O$,
respectively.

For each group $G_g$, construct a bipartite
multigraph on disjoint copies of $S$ and $O$.
For every client $j\in G_g$, add an edge
joining $\sigma_S(j)$ and $\sigma_O(j)$.

Since both assignments respect the capacity
$b_g$, the multigraph has maximum degree
at most $b_g$. By K\"onig's edge-coloring theorem for
bipartite multigraphs, its edges can be
properly colored using at most $b_g$ colors.
Each color class is a matching, since no
two edges of the same color share an endpoint.

The corresponding client sets therefore
partition $G_g$ into at most $b_g$
refined groups.
Under both $\sigma_S$ and $\sigma_O$,
every facility serves at most one client
from each refined group.

Applying this construction independently
to every original group gives a refinement
of the partition of $V_C$, where all
refined groups have capacity one.

Consider the facility-location instance
defined by these refined groups, with
the same metric and facility-opening costs.
Let $\widetilde A(X)$ denote the minimum
connection cost for a facility set $X$
in this refined instance, taking
$\widetilde A(X)=+\infty$ if $X$ is infeasible.

Every assignment feasible for the refined
instance also respects the original
group capacities. Hence
\[
    \widetilde A(X)\geq A(X)
    \qquad(X\subseteq V_F).
\]

Moreover, $\sigma_S$ and $\sigma_O$
remain feasible under the refined
capacity-one constraints.
Since they were minimum-cost assignments
in the original instance, we have
\[
    \widetilde A(S)=A(S)=C,
    \qquad
    \widetilde A(O)=A(O)=C^*.
\]

We first show that $S$ remains locally
optimal in the refined instance.
Suppose, for contradiction, that some
add, delete, or single-swap move from
$S$ to $X$ improves the cost in the
refined instance. Then
\[
\begin{aligned}
    F(X)+A(X)
    &\leq F(X)+\widetilde A(X)\\
    &<F(S)+\widetilde A(S)\\
    &=F(S)+A(S),
\end{aligned}
\]
contradicting the local optimality of
$S$ in the original instance.

Thus, $S$ remains locally optimal
in the refined instance.

Similarly, $O$ remains optimal in
the refined instance.
Indeed, for every facility set
$X\subseteq V_F$,
\[
\begin{aligned}
    F(X)+\widetilde A(X)
    &\geq F(X)+A(X)\\
    &\geq F(O)+A(O)\\
    &=F(O)+\widetilde A(O),
\end{aligned}
\]
where the second inequality follows
from the optimality of $O$ in the
original instance.

We can therefore apply
Lemma~\ref{lem:capacity-one-nonuniform}
to the refined instance, obtaining
\[
    C+F\leq 15(C^*+F^*).
\]

Finally, the standard improvement-threshold
argument discussed in
Subsection~\ref{subsec:our-results-and-techniques}
gives a polynomial-time local search
algorithm with an arbitrarily small
additional loss $\epsilon$ in the
approximation ratio.
This proves the claimed
$(15+\epsilon)$-approximation guarantee.
\end{proof}

\subsection{Challenges for Arbitrary Matroids
and Non-uniform Opening Costs}
\label{subsec:arbitrary-matroid-challenges}

It is natural to ask whether the same
local search algorithm gives a
constant-factor approximation for
\flmi with an arbitrary common client
matroid and non-uniform facility-opening costs.

Our analysis for partition matroids
relies on the fact that their independence
constraints decompose across groups.
This allows us to refine each group
into capacity-one groups and analyze
the client reassignments separately.

This approach does not directly extend
to an arbitrary matroid.
In a general matroid, two sets of clients
may each be independent while their union
is dependent.
Therefore, constructing feasible
reassignments for different subsets
of clients separately does not necessarily
yield a feasible assignment when those
reassignments are combined.

In particular, the test-swap argument
from Subsection~\ref{subsec:capacity-one}
relies on treating the affected groups
independently.
Without the partition-matroid structure,
we cannot directly apply the same
reassignment construction to control
the connection-cost increase of the swaps.

However, this difficulty concerns
our analysis rather than the local
search algorithm itself.
The failure of the preceding argument
to extend to arbitrary matroids does
not rule out a constant locality gap.
We propose the following conjecture.

\begin{conjecture}
\label{conj:general-matroid-nonuniform}
There exists an constant
$\alpha\geq 1$ such that, for every
instance of \flmi with an arbitrary
common client matroid and nonnegative
facility-opening costs, every solution
$S$ that is locally optimal with
respect to add, delete, and single-swap
moves satisfies
\[
    \cost(S)\leq\alpha\cost(O),
\]
where $O$ is an optimal solution.
\end{conjecture}

We note that our proofs of Lemmas \ref{lem:connection-bound} and \ref{lem:NOPfacility-bound} hold in this general setting, so to prove Conjecture \ref{conj:general-matroid-nonuniform} it suffices to show $F(P)$ is bounded by $O(OPT)$ as we did in Sections \ref{sec:gen-matroid-unif-cost} and \ref{sec:group-non-uniform} to prove Theorems \ref{thm:general-matroid-approximation} and \ref{thm:partition-matroid-nonuniform}, respectively.

\printbibliography
\appendix

\section{AI Methodology}

We used ChatGPT (OpenAI) to help
identify the potential function $h(x)$ used in
Section~\ref{subsec:Bounded-congestion mapping}, as well as to improve
the grammar, clarity, and readability of the manuscript.
Its mathematical contribution was limited to suggesting this potential;
the subsequent mathematical analysis, derivation of the bounds, and proofs
were developed independently by the authors.
The authors verified the correctness and originality of all content,
including references, and take full responsibility for the final manuscript.

\section{Hardness Beyond the Structural Assumptions}
\label{app:hardness}

\subsection{Overlapping Group Constraints}
\label{app:overlapping-groups-hardness}

Consider a variant of $\flmi$ with group constraints where the groups need not necessarily partition the set of clients.
That is, the constraints dictating what is feasible are given by possibly overlapping groups of clients and
we can take a most $b_g$ clients from each group. Essentially, this variant captures graph colouring.

\begin{theorem}
For every fixed $\epsilon>0$, facility location with
unrestricted overlapping group constraints admits no
polynomial-time $n^{1-\epsilon}$-approximation unless
$P=NP$, where $n$ is the number of clients.
\end{theorem}

\begin{proof}
We give an approximation-preserving reduction from
Graph Coloring.

Let $G=(V,E)$ be an undirected graph with $n=|V|$.
Construct a facility-location instance as follows.
Create one client for each vertex $v\in V$ and one
group $\{u,v\}$ for each edge $uv\in E$.
Create $n$ facilities and colocate all clients and
facilities at a single location, so that every
connection cost is zero. Assign the same opening
cost $n$ to every facility.

Under the group constraints, each facility may
serve at most one client from each group.
Therefore, the clients assigned to any facility
must form an independent set in $G$.

We first show that $G$ admits a proper coloring
with at most $k$ colors if and only if the
constructed instance admits a feasible assignment
using at most $k$ facilities.

Suppose that $G$ has a proper coloring with at most
$k$ colors. Open one facility for each color class
and assign every client to the facility corresponding
to its color. Since each color class is an
independent set, no facility receives both endpoints
of any edge. Hence all group constraints are
satisfied.

Conversely, suppose that the constructed instance
admits a feasible assignment using at most $k$
facilities. Assign to each vertex the color
corresponding to the facility serving its client.
For every edge $uv\in E$, the clients $u$ and $v$
belong to the same group and therefore cannot be
assigned to the same facility. Thus, adjacent
vertices receive different colors, yielding a
proper coloring with at most $k$ colors.

It follows that the minimum number of facilities
required by a feasible solution is exactly
$\chi(G)$, the chromatic number of $G$.
Since all connection costs are zero and every
facility has opening cost $n$, the optimal
facility-location cost is
\[
    \mathrm{OPT}_{\mathrm{FL}}=n\chi(G).
\]

Moreover, any feasible facility-location solution
opening $k$ facilities has cost $nk$ and induces
a proper coloring using at most $k$ colors.
Therefore, if a polynomial-time algorithm achieved
approximation ratio $\rho(n)$ for the constructed
facility-location instance, it would yield a
polynomial-time coloring algorithm satisfying
\[
    k \le \frac{\rho(n)\mathrm{OPT}_{\mathrm{FL}}}{n}
      = \rho(n)\chi(G).
\]
The reduction is polynomial-time and preserves
approximation ratios.

By the $n^{1-\epsilon}$-inapproximability of
Chromatic Number for every fixed $\epsilon>0$
\cite{FeigeKilian1998,Hastad1999,zuckerman2007}, the claimed hardness follows
unless $P=NP$.
\end{proof}

\subsection{Facility-Specific Client Matroids}
\label{app:facility-specific-hardness}

In this section, we show that if instead of a common matroid $\mathcal M$ describing feasibility at a facility, each possible facility $i \in V_F$ has its own matroid $\mathcal M_i$ over the clients $V_C$
describing the subsets of clients that can be assigned to $i$ in a feasible solution. Essentially, we show this generalization captures the difficulty of approximating set cover.

\begin{theorem}
For every fixed $0<\eta<1$, facility location
with facility-specific client matroids admits no
polynomial-time $(1-\eta)\ln n$-approximation
unless $P=NP$, even when all opening costs are
uniform and all connection costs are zero,
where $n$ is the number of clients.
\end{theorem}

\begin{proof}
We give an approximation-preserving reduction
from Set Cover.

Consider a Set Cover instance with universe
$U$, where $|U|=n$, and a collection of sets
$S_1,\ldots,S_m\subseteq U$.
We may assume that these sets cover $U$.
Let $\tau$ denote the minimum number of sets
whose union is $U$.

Construct a facility-location instance with
one client for each element of $U$ and one
facility $i$ for each set $S_i$.
Colocate all clients and facilities at a single
location, so that every connection cost is zero.
Give every facility the same opening cost one.

For each facility $i$, define a client matroid
$\mathcal M_i=(U,\mathcal I_i)$ with
\[
    \mathcal I_i
    =\{I\subseteq U:I\subseteq S_i\}
    =2^{S_i}.
\]
This is the matroid that is free on $S_i$
and has every element of $U\setminus S_i$
as a loop. Consequently, a set of clients
can be assigned to facility $i$ if and only
if it is a subset of $S_i$.

We show that the Set Cover instance admits
a cover of size at most $k$ if and only if
the constructed facility-location instance
admits a feasible assignment using at most
$k$ facilities.

Suppose that there exists a set cover
$S_{i_1},\ldots,S_{i_k}$.
Open the corresponding facilities and assign
each client $u\in U$ to any opened facility
$i_j$ satisfying $u\in S_{i_j}$.
Such a facility exists because the selected
sets cover $U$. Moreover, every facility
$i_j$ receives only clients belonging to
$S_{i_j}$, so its assigned client set is
independent in $\mathcal M_{i_j}$.
Thus the assignment is feasible.

Conversely, suppose that a feasible
facility-location solution opens at most
$k$ facilities, indexed by a set $F$.
Every client $u\in U$ must be assigned to
some facility $i\in F$. By the definition
of $\mathcal M_i$, this is possible only
if $u\in S_i$. Therefore,
\[
    U=\bigcup_{i\in F}S_i,
\]
and the corresponding sets form a set cover
of size at most $k$.

Hence the minimum number of open facilities
is exactly $\tau$. Since all connection
costs are zero and each facility has opening
cost one, the optimal facility-location
cost satisfies
\[
    \mathrm{OPT}_{\mathrm{FL}}=\tau.
\]

Furthermore, any feasible solution opening
$k$ facilities has cost $k$ and directly
induces a set cover of size at most $k$.
Therefore, a polynomial-time
$\rho(n)$-approximation for facility location
would yield a polynomial-time Set Cover
algorithm satisfying
\[
    k\le \rho(n)\mathrm{OPT}_{\mathrm{FL}}
      =\rho(n)\tau.
\]
The construction is polynomial-time and
preserves approximation ratios.

By the $(1-\eta)\ln n$-inapproximability
of Set Cover for every fixed $0<\eta<1$
\cite{Feige1998, DinurSteurerSetCover2014}, the claimed
hardness follows unless $P=NP$.
\end{proof}

\end{document}